\documentclass[amsmath,amssymb,twocolumn,pre]{revtex4-1}
\usepackage{bbm}
\usepackage{amssymb}

\usepackage{epsfig}
\usepackage{graphicx}
\usepackage{mathrsfs}
\usepackage{amsfonts}
\usepackage{color}

\begin{document}

\title{Nonsteady dynamics at the dynamic depinning transition in two-dimensional Gaussian random-field Ising model }

\author{Xiaohui Qian}
\author{Gaotian Yu}
\author{Nengji Zhou}
\email[Corresponding author:~]{zhounengji@hznu.edu.cn}
\date{\today}
\affiliation{School of Physics, Hangzhou Normal University, Hangzhou 311121, China
}

\begin{abstract}
With large-scale Monte Carlo simulations, we investigate the nonsteady relaxation at the dynamic depinning transition in two-dimensional Gaussian random-field Ising model. The dynamic scaling behavior is carefully analyzed, and the transition fields as well as static and dynamic exponents are accurately determined based on the short-time dynamic scaling form. Different from the usual assumption, two distinguished growth processes of spatial correlation lengths for the velocity and height of the domain wall are found. The universality class of the depinning transition is thus established which significantly differs from that of the QEW equation, but agrees with that of the recent experiment as well as other simulations works. Under the influence of the mesoscopic time regime, the crossover from the second-order phase transition to the first-order one is confirmed in the weak disorder regime, yielding an abnormal disorder-dependent nature of the criticality.
\end{abstract}

\maketitle

\section{Introduction}\label{sec:intro}
Magnetic domain-wall motions in random media have been extensively studied over the last two decades because of their potential application to memory, logic, and neuromorphic devices \cite{all05,hay08,par08,zha12,luo20}. Intense experimental and theoretical efforts have been devoted to manipulating domain walls in response to various external forces, including magnetic fields, electric fields, spin-orbit torques, spin waves, etc \cite{lee11,ger14,ram15,kim22,jac22,liu22,lan22}. As shown in recent studies \cite{bol04,im09,dou06,gor14,ska22}, there exists the so-called ``depinning transition'' separating the pinned state from the moving phase with a nonzero steady-state velocity due to the competition between the driving force and disorder at zero temperature. When the temperature is low but nonvanishing, the sharp transition is softened and thermally activated creep motion occurs even when the driving force is far below the depinning threshold \cite{lem98,dut16,jin21}. Under a periodic harmonic driving, domain-wall motions exhibit four different states, i.e., relaxation, creep, sliding, and switching. Rich and novel transitions have been found among them, e.g., the relaxation-to-creep transition \cite{che02,bra05,zho10,zho14}. However, the understanding of such dynamic transitions is still limited, especially for the growth of the correlation length.

Besides domain-wall motions in ferromagnetic and ferroelectric materials \cite{ban14,xi15}, in fact, the dynamic depinning transition is also the focus of researches on contact lines in wetting \cite{du14}, charge-density waves \cite{bra04}, liquid invasion in porous media \cite{ros07}, vortices in type-II superconductors \cite{nat00,oku12}, and dislocation dynamics in crystal plasticity \cite{ova15}. However, experimental studies of depinning transitions are likely to be challenging, since a very low temperature is required to minimize the thermal rounding of the velocity-force characteristics \cite{luo07, bus08,gor14}. Very recently, depinning behaviors have been experimental evidenced for a wide set of materials, and a universal scaling function accounting for both drive and thermal effects has been given \cite{par17}. Moreover, direct experimental determination of critical exponents has been performed for the depinning transition, yielding the velocity exponent $\beta=0.30(3)$ and correlation-length exponent $\nu=1.3(3)$ \cite{alb21a}.

Theoretically, a phenomenological model called as Edwards-Wilkinson equation with quenched disorder (QEW) is commonly used to study the depinning transition wherein the domain wall is effectively described by a single-valued elastic string \cite{due05,kol06a,kim06}. However, simulations of the QEW equation consistently yield the ``superrough'' behavior at the depinning transition with the roughness exponent $\zeta=1.25(1) > 1$ which goes beyond the simple description of the domain wall. Besides, the velocity exponent $\beta=0.245(6)$ \cite{fer13} has been determined recently, smaller than the experimental result. It suggests that detailed microscopic structures and interactions of real materials should be concerned, which allows a closer comparison with experiments.

With Monte Carlo simulations as well as micromagnetic simulations based on the stochastic Landau-Lifshitz-Gilbert (LLG) equation, depinning transitions in microscopic lattice models have been well studied in recent years \cite{qin12a,he16,jin19,zho09}. Both static and dynamic critical exponents have been accurately estimated, and results show that they are not in the universality class of the QEW equation. However, the issue regarding whether these microscopic details lead to a new universality class remains unaddressed, and hyperscaling relations have yet to receive full support \cite{zho16}. Very recently, an unusual disorder-dependent nature of criticality has been revealed by simplified LLG simulations based on Lagrangian dynamics equation, associated with the proliferation of Bloch lines within the domain wall at strong disorder \cite{ska22}. It is worth noting that the quenched disorder with Gaussian distribution and a nonzero correlation length was different from those used in lattice simulations that employed uniformly distributed random fields which were simple and bounded. It begs the question of what happens when the disorder realization shifts to the unbounded Gaussian distribution. Further research is necessary to determine the universality of the transition.

Thanks to the short-time critical dynamics (STCD) method originally developed by Janssen and collaborators \cite{jan89}, one can obtain equilibrium critical exponents from the scaling analysis of the relaxation dynamics far away from the equilibrium, without being hindered by critical slowdown \cite{zhe98,zho08,alb11,zho13}. Recently, this efficient method has been employed to tackle the dynamic depinning transition in both the QEW equation \cite{kim06,kol06,fer13} and lattice models \cite{jin19,zho09,zho16}. Compared with those from steady-state motions, more precise results of the critical exponents along with the transition fields can be extracted from the nonsteady relaxation owing to large-scale simulations. Besides, it is known that the growing correlation length plays a crucial role in the critical dynamics. For the dynamic transition, however, there are two kinds of spatial correlation lengths for the velocity and height of the domain wall, respectively. The former is related to the phase transition because the velocity is the order parameter. While the latter governs the roughening process of the domain wall, but has always been mistaken for the velocity correlation length \cite{kol06,zho09,zho16,fer13}.

The aim of this paper is to identify the universality of the depinning transition in two-dimensional Gaussian random-field Ising model. By extensive simulations based on the extended Monte Carlo algorithm, nonsteady dynamics at the dynamic depinning transition is carefully examined, and critical exponents as well as the transition fields are accurately determined by the STCD method, in comparison with those of the QEW equation and recent experiment. In addition, the growths of the correlation lengths extracted from the fluctuations and correlations are analyzed, and the influence of the disorder on the depinning transition is investigated. In Sec.~\ref{sec:mod}, the models and scaling analysis are described, and in Sec.~\ref{sec:simulation}, the numerical results are presented. Finally, Sec.~\ref{sec:conclusion} includes the conclusion.

\section{Model and scaling analysis}\label{sec:mod}
\subsection{Model}
The Hamiltonian of random-field Ising model with a driving field is given by
\begin{equation}
\label{Hamiltonian}
\mathcal H=-J\sum_{\left\langle ij\right\rangle}S_iS_j-\sum_{i}(h_i+H)S_i,
\end{equation}
where $S_i=\pm1$ is the Ising spin at site $i$ of a rectangle lattice $2L\times L$, $\left\langle ij\right\rangle$ denotes a summation over nearest neighbors, $H$ represents a homogeneous driving field, and $h_i$ stands for a random field of Gaussian distribution with zero mean and standard deviation $\sigma$. In this work, Monte Carlo simulations are performed at zero temperature with the time up to $t_{\rm max}=20000$ Monte Carlo Step (MCS), and the number of samples for average is about $20000$. Statistical errors are estimated by dividing the total samples into two subgroups. If the fluctuation in the time direction is comparable with or larger than the statistical error, it will be taken into account. The initial condition is a semi-ordered state with a flatted domain wall in the $y$ direction where the spins are positive on the left side and negative on the right side. Antiperiodic and periodic boundary conditions are used in the $x$ and $y$ directions, respectively. An extended Monte Carlo algorithm is adopted to improve the efficiency of simulations \cite{zho16}.  Maintaining the bulk spin invariant throughout the simulation is crucial in this study, as we only focus on the depinning transition with a single domain interface. Thus, domain nucleation in the bulk has been precluded entirely during the simulation, even under strong disorder conditions.

Denoting a spin at site $(x,y)$ by $S_{xy}(t)$, the height function of the domain wall is given by
\begin{equation}
\label{height function definition}
h(y,t)=\frac{L_x}{2}\left[m(y,t)+1\right],
\end{equation}
where $m(y,t)$ is the line magnetization defined as
\begin{equation}
\label{line magnetization}
m(y,t)=\frac{1}{L_x}\left[\sum_{x=1}^{L_x}S_{xy}(t)\right].
\end{equation}
Local velocity can be calculated from the height function,
\begin{equation}
\label{local velocity function definition}
v(y,t)=\frac{dh(y,t)}{dt}.
\end{equation}
The average velocity $v(t)$ of the domain wall is then obtained,
\begin{equation}
\label{velocity function definition}
v(t)=\left\langle v(y,t) \right\rangle,
\end{equation}
where $\left\langle\dots\right\rangle$ includes both the statistical average over samples and in the $y$ direction. Moreover, local and global velocity fluctuations are introduced,
\begin{equation}
\label{line susceptibility of velocity}
\omega^2_v(t)=\left\langle v(y,t)^2\right\rangle-v(t)^2,
\end{equation}
and
\begin{equation}
\label{planar susceptibility of velocity}
v^{(2)}(t)=\left\langle\left[ \frac{1}{L}\sum_{y=1}^{L}v(y,t)\right]^2\right\rangle-v(t)^2.
\end{equation}
Considering that $\omega^2_v(t)$ and $v^{(2)}(t)$ also reflect the line and planar susceptibility of the velocity, respectively, the second-order cumulant of the order parameter is derived,
\begin{equation}
\label{Fv}
F_v(t) = \frac{v^{(2)}(t)}{\omega^2_v(t)}.
\end{equation}
Another important observable is the velocity correlation function,
\begin{equation}
\label{velocity correlation function}
C_v(r,t)=\left\langle v(y+r,t)v(y,t)\right\rangle-v(t)^2,
\end{equation}
which describes the pure correlation of the velocity in the $y$ direction.

Besides, the roughening process of domain walls is also investigated by means of the roughness function $\omega_h^2(t)$ and correlation function $C_h(r,t)$,
\begin{equation}
\label{roughness function definition}
\omega_h^2(t)=\left\langle h(y,t)^2\right\rangle-\left\langle h(y,t)\right\rangle^2,
\end{equation}
and
\begin{equation}
\label{height correlation function}
C_h(r,t)=\left\langle h(y+r,t)h(y,t)\right\rangle-\left\langle h(y,t)\right\rangle^2.
\end{equation}
Similar to the cumulant $F_v(t)$, the function $F_h(t)$ is introduced as the ratio of the planar and line susceptibilities,
\begin{equation}
\label{Fh}
F_h(t)=\frac{M^{(2)}(t)-M(t)^2}{\omega^2_h(t)},
\end{equation}
where $M(t)$ and $M^{(2)}(t)$ represent the magnetization and its second moment, respectively.

\subsection{Scaling analysis}
The dynamic scaling form of the depinning transition can be derived by the STCD arguments \cite{zhe98,zho09},
\begin{equation}
\label{scaling 1}
v(t,\tau,L)=b^{-\beta/\nu}G(b^{-1}\xi(t),b^{1/\nu}\tau,b^{-1}L),
\end{equation}
where $b$ is an arbitrary rescaling factor, $\beta$ and $\nu$ correspond to the static exponents for the velocity and correlation length, respectively, $\tau=|H-H_c|$ denotes the deviation from the critical field, and $\xi(t)$ represents the spatial correlation length. The scaling theory holds in the macroscopic regime $t>t_{\rm mic}$ where $t_{\rm mic}$ denotes a microscopic time scale. Generally speaking, the value of $t_{\rm mic}$ is not universal, and relies on microscopic details of the dynamic systems. For the simple Ising model with the nearest-neighbor interactions, $t_{\rm mic}$ is rather short about $20$ MCS \cite{zho08}. Setting $b=\xi(t)$, the scaling form can be simplified to
\begin{equation}
\label{scaling 2}
v(t,\tau)=\xi(t)^{-\beta/\nu}G(\xi(t)^{1/\nu}\tau),
\end{equation}
in the short-time scaling regime with $\xi(t)\ll L$. Thus, a power-law behavior of the velocity is expected at the transition point $\tau=0$,
\begin{equation}
\label{scaling 3}
v(t)\sim~\xi(t)^{-\beta/\nu}.
\end{equation}
From Eq.~(\ref{scaling 2}), the time derivative of $v(t,\tau)$ is simply derived to determine the exponent $\nu$,
\begin{equation}
\label{scaling 4}
\frac{\partial\ln v(t,\tau)}{\partial \tau}|_{\tau=0}\sim \xi(t)^{1/\nu}.
\end{equation}

At the second-order transition, it is believed that the spatial correlation length grows as
\begin{equation}
\label{scaling 5}
\xi(t)\sim~t^{1/z},
\end{equation}
where $z$ is the dynamic critical exponent. Thus, one can determine the values of transition point $H_c$ and critical exponent $\beta/\nu z$ by searching for the best power-law behavior of $v(t)$, which is a standard procedure in the STCD method. Due to the existence of strong correction in $\xi(t)$, however, the above procedure suffers. Here we present an alternative way by investigating the scaling behavior of $v(\xi)$. With Eq.~(\ref{scaling 3}), the transition point as well as the critical exponent can be accurately estimated.

Now the problem is how to measure the intrinsic length scale $\xi_v(t)$ at the depinning transition. Similar to that of the order parameter, the scaling form of the velocity fluctuation is derived,
\begin{equation}
\label{scaling 6}
v^{(2)}(t)=\omega ^2_v(t)G(\xi_v(t)/L).
\end{equation}
For a sufficiently large lattice $\xi_v(t)\ll L$, the scaling behavior of the cumulant $F_v(t)$ is expected by using the finite-size scaling analysis,
\begin{equation}
\label{scaling 7}
F_v(t)= G(\xi_v(t)/L)\sim(\xi_v(t)/L)^d,
\end{equation}
where $d=1$ is the dimension of the domain wall. Therefore, the correlation length $\xi(t)$ in Eq.~(\ref{scaling 3}) can be replaced by the cumulant $F_v(t)$ to locate the transition point $H_c$. Moreover, the dynamic exponent $z$ and the correction to scaling of $\xi_v(t)$ are extracted from the time dependence of the cumulant $F_v(t)$. Further analyses give power-law behaviors of local and global velocity fluctuations at the transition $H_c$,
\begin{eqnarray}
\label{scaling 8}
\omega^2_v(t) & \sim & t^{d_1/z-2\beta/\nu z-\delta}, \nonumber \\
v^{(2)}(t)& \sim & t^{d_2/z-2\beta/\nu z-\delta}/L,
\end{eqnarray}
where $\delta$ is a critical exponent characterizing the dynamic effect of overhangs and islands \cite{zho10a}, and $d_1=1$ and $d_2=2$ correspond to the spatial dimensions of the line and planar velocity susceptibilities, respectively.

Besides, the correlation length $\xi_v(t)$ can also be measured by the velocity correlation function $C_v(r,t)$ defined in Eq.~(\ref{velocity correlation function}) with the scaling form
\begin{equation}
\label{scaling 9}
C_v(r,t)=\omega^2_v(t)\widetilde C_v(r/\xi_v).
\end{equation}
A power-law behavior of the scaling function $\widetilde C_v(s)$ is assumed based on general scaling arguments,
\begin{equation}
\label{scaling 10}
\widetilde C_v(s)\sim s^{-2\beta/\nu-\delta z},
\end{equation}
when the ratio $s=r/\xi_v(t)\leq 1$.

Similarly, the correlation length related to the height $\xi_h(t)$ can be obtained from the susceptibility ratio $F_h(t)$ and correlation function $C_h(r,t)$ with the scaling forms
\begin{equation}
\label{scaling 11}
F_h(t)= G(\xi_h(t)/L)\sim\xi_h(t)/L,
\end{equation}
and
\begin{equation}
\label{scaling 12}
C_h(r,t)=\omega^2_h(t)\widetilde C_h(r/\xi_h),
\end{equation}
respectively. However, the scaling function $\widetilde C_h(s=r/\xi_h)$ decays exponentially,
\begin{equation}
\label{scaling 13}
\widetilde C_h(s)\sim \exp[-s^{2\zeta _{\rm loc}}],
\end{equation}
much faster than $\widetilde C_v(s)$, where $\zeta _{\rm loc}$ represents the local roughness exponent.
With the correlation length $\xi_h(t)$ at hand, the roughness exponent $\zeta$ can be measured by
\begin{equation}
\label{scaling 14}
\omega_h^2(t)\sim\left[ \xi_h(t)\right]^{2\zeta}.
\end{equation}

Finally, corrections to scaling are also considered by extending the fitting to early times, though power-law behaviors of $v(t)$, $v^{(2)}(t)$, $\omega^2_v(t)$, $\widetilde C_v(s)$, and $\omega_h^2(t)$ are expected at the critical point. Usually, a power-law correction form is adopted,
\begin{equation}
\label{correction}
y=ax^b(1+c/x),
\end{equation}
where the fitting parameter $b$ stands for the critical exponent.
\section{NUMERICAL SIMULATIONS}\label{sec:simulation}
\subsection{Determination of the transition point and critical exponents}
\begin{figure*}[htb]
\centering
\epsfysize=8cm \epsfclipoff \fboxsep=0pt
\setlength{\unitlength}{1.cm}
\begin{picture}(9,8)(0,0)
\put(0.0,0.0){{\epsffile{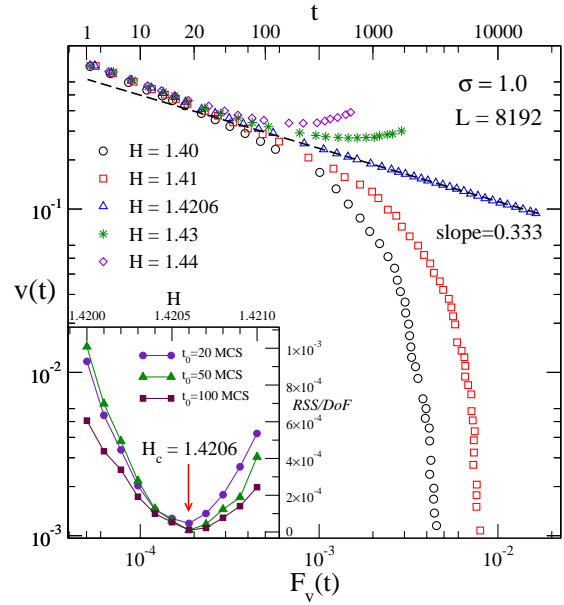}}}
\end{picture}
\caption{ The velocity of the domain wall $v(t)$ is displayed as a function of $F_v$(t) which is the second-order cumulant of the order parameter for different driving fields $H$ on a log-log scale. The upper $t$ scale corresponds to that of $F_v(t)$, and the dashed line represents a power-law fit. In the inset, the fitting error within a very narrow $H$ regime is shown for different waiting time $t_0$, and the transition point $H_c$ is indicated by the arrow.}
\label{f1}
\end{figure*}

With extensive Monte Carlo simulations, we investigate nonsteady dynamics at the depinning phase transition under Gaussian noise, taking the setting of the lattice size $L=8192$ and disorder strength $\sigma=1.0$. In Fig.~\ref{f1}, the domain-wall velocity $v(t)$ against the cumulant $F_v(t)$ is plotted for different driving fields $H$. It drops rapidly for a smaller $H$, and approaches a constant for a larger $H$. One then locates the transition field $H_c\approx1.42$ where the curve exhibits a power-law behavior starting from the rather early time around $20$ MCS. To obtain a more accurate value of the critical point, the fitting error $RSS/DoF$ is carefully examined within a very narrow window $H\in[1.420, 1.421]$, as shown in the inset. By judging the position of the minimal $RSS/DoF$, one obtains $H_c=1.4206(1)$ for three different waiting times $t_0= 20, 50$, and $100$ MCS, showing the result is robust and accurate. Without losing generality, all critical exponents in the following are measured with $t_0=100$ MCS. For example, the exponent $\beta/\nu=0.333(3)$ in Eq.~(\ref{scaling 3}) can be estimated from the slope of the curve in such a time interval.

\begin{figure*}[htb]
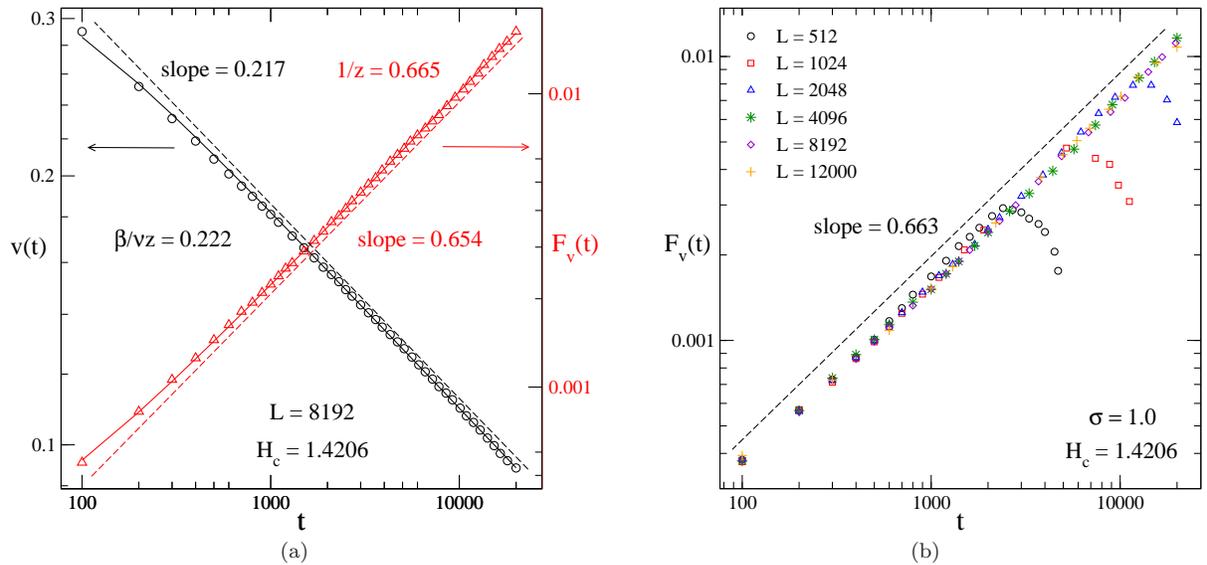

\centering
\epsfysize=7cm \epsfclipoff \fboxsep=0pt
\setlength{\unitlength}{1.cm}
\begin{picture}(10,7)(0,0)
\put(-3.0, 0.0){{\epsffile{vt-Fvt.eps}}}\epsfysize=7cm
\put(5.7, 0.0){{\epsffile{finite_effect.eps}}}
\end{picture}

\hspace{0.0cm}\footnotesize{(a)}\hspace{8.0cm}\footnotesize{(b)}
\caption{(a) The time-dependent functions $v(t)$ and $F_v$(t) are given with open circles and open triangles, respectively, at the critical point $H_c=1.4206$ for the disorder strength $\sigma=1.0$ and lattice size $L=8192$ on a log-log scale. The right and left arrows indicate the $y$ coordinates of $v(t)$ and $F_v$(t), respectively. (b) The cumulant $F_v(t)$ is displayed for different lattice sizes $L$ on a log-log scale. Dashed lines represent power-law fits, and solid lines in (a) show power-law fits with the correction. All the curves in (b) are rescaled by the factor $L/L'$ where $L'=12000$ is fixed.   }
\label{f2}
\end{figure*}

In Fig.~\ref{f2}(a), the time dependence of the velocity $v(t)$ and second-order cumulant $F_v(t)$ at $H_c$ are investigated with open circles and open triangles, respectively. With the usual scaling forms in Eqs.~(\ref{scaling 3}) and (\ref{scaling 5}), the exponents $\beta/\nu z=0.217(4)$ and $1/z=0.654(8)$ are measured from the slopes of dashed lines. However, a significant departure from the power-law behavior is found in early times, suggesting the existence of corrections to scaling.  With the form in Eq.~(\ref{correction}), we refine the exponents $\beta/\nu z=0.222(4)$ and $1/z=0.665(8)$ from which one can calculate the dynamic exponent $z=1.50(2)$, in good agreement with that of QEW equation \cite{fer13,kol06}.

Moreover, the finite-size effect described by $G(\xi_v(t)/L)$ in Eq.~(\ref{scaling 7}) is studied in Fig.~\ref{f2}(b) on a log-log scale. The cumulant $F_v(t)$ is displayed for different lattice sizes $L=512,1024,2048,4096,8192$, and $12000$. Subsequently, we fix a lattice size, e.g., $L'=12000$, and change the scale of $F_v(t)$ of another $L$ to $F_v(t)(L'/L)^{-1}$. All the data of different $L$ then collapses to the master curve with the slope $0.663(3)$, confirming the finite-size dependence $L^{-1}$ in Eq.~(\ref{scaling 7}) when $\xi_v(t) \ll L$. The dynamic exponent $z=1.51(1)$ is estimated again, the same as that in Fig.~\ref{f2}(a) within the error bar. The curves of $L = 8192$ and $12000$ are nicely overlapped up to the time $t_{\max}=20000$ MCS, suggesting that the finite-size effect of the correlation length $\xi_v(t)$ is already negligibly small.

\begin{figure*}[htb]
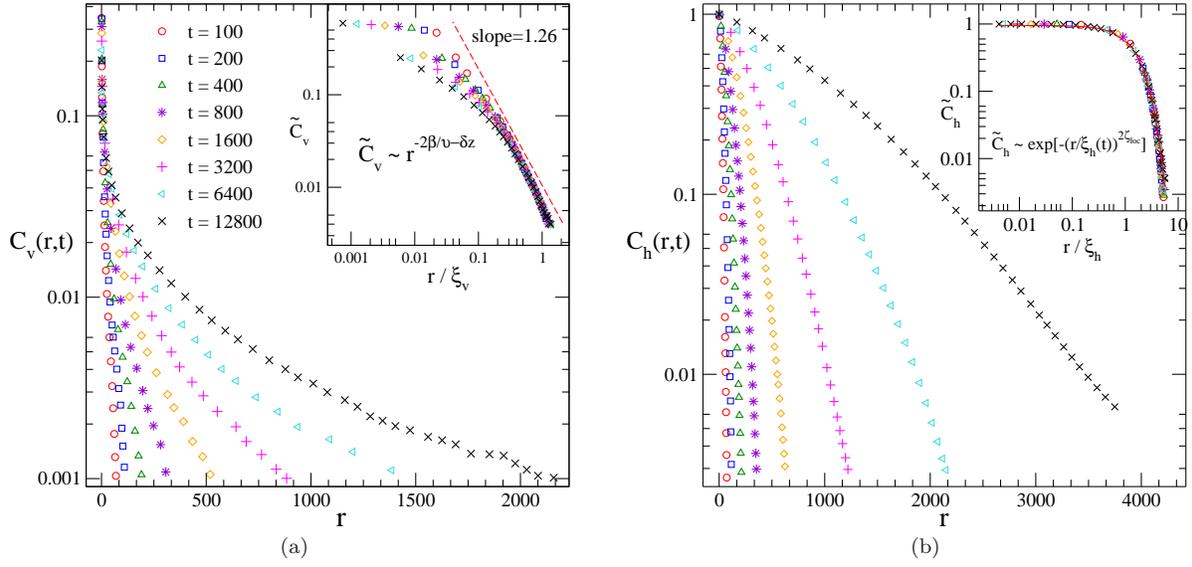

\centering
\epsfysize=7cm \epsfclipoff \fboxsep=0pt
\setlength{\unitlength}{1.cm}
\begin{picture}(10,7)(0,0)
\put(-3.0, 0.0){{\epsffile{Cv.eps}}}\epsfysize=7cm
\put(5.2, 0.0){{\epsffile{Ch.eps}}}
\end{picture}

\hspace{0.0cm}\footnotesize{(a)}\hspace{8.0cm}\footnotesize{(b)}
\caption{ The velocity correlation function $C_v (r,t)$ in (a) and correlation of height function $C_h (r,t)$ in (b) are plotted for different time $t$ at the disorder strength $\sigma=1.0$ and critical driving field $H_c=1.4206$ on a linear-log scale. In the insets, scaling functions $\widetilde C_v(r,t)=C_v/\omega^2_v$ and $\widetilde C_h(r,t)=C_h/\omega^2_h$ are shown with the variable $r/\xi(t)$ on a log-log scale, and data collapses are demonstrated. The dashed and solid lines represent the fittings with the power-law and exponential forms, respectively.}
\label{f3}
\end{figure*}

A more direct way to extract the correlation length is to study the velocity correlation function $C_v(r,t)$ and the correlation of the height function $C_h(r,t)$. As shown in Fig.~\ref{f3}, numerical data at different time $t$ can collapse to the curve at $t'=12800$ MCS by rescaling $r$ to $[\xi(t')/\xi(t)]r$ and $C(r,t)$ to $[\omega^2(t')/\omega^2(t)]C(r,t)$ with an adjustable parameter $\xi(t)$ as input. By adopting such data collapse technique, we determine the correlation lengths $\xi_v(t)$ and $\xi_h(t)$ in the subgraphs (a) and (b), respectively. Insets show the scaling behaviors of $\widetilde C_v(r/\xi_v)$ and $\widetilde C_h(r/\xi_h)$ defined in Eqs.~(\ref{scaling 9}) and (\ref{scaling 12}). In the left panel, a power-law decay of  $\widetilde C_v(r/\xi_v)$ is found with the slope $1.26(5)$, yielding the overhang exponent $\delta=(1.26-2\beta/\nu)/z=0.40(2)$. It is inferred that overhangs and islands which naturally develop nearby the domain wall essentially affect the dynamic scaling behavior of the velocity correlation. Furthermore, the scaling function $\widetilde C_v(r/\xi_v)$ is not universal for a small distance $r \ll \xi_v(t)$, consistent with steady-state results reported in Ref.~\cite{due05}.

Unlike $\widetilde C_v(r/\xi_v)$, the scaling form of the height correlation function $\widetilde C_h(r/\xi_h)$ shown in the right panel is universal over the entire range of $r/\xi_h$, indicating that the contribution of the overhangs and islands to the roughening process of the domain wall becomes suppressed due to the definition of the domain-wall height in Eq.~(\ref{height function definition}). An exponential decay of $\widetilde C_h(r/\xi_h)$ rather than a power-law one is found, and the local roughness exponent $2\zeta_{loc}=1.38(5)$ is measured from the fitting to numerical data based on  Eq.~(\ref{scaling 13}). The result $\zeta_{loc}=0.69$ is comparable with the experimental results in the ultrathin Pt/Co/Pt films \cite{lem98,met07,bur21}.

\begin{figure*}[htb]
\centering
\epsfysize=8cm \epsfclipoff \fboxsep=0pt
\setlength{\unitlength}{1.cm}
\begin{picture}(9,8)(0,0)
\put(0.0,0.0){{\epsffile{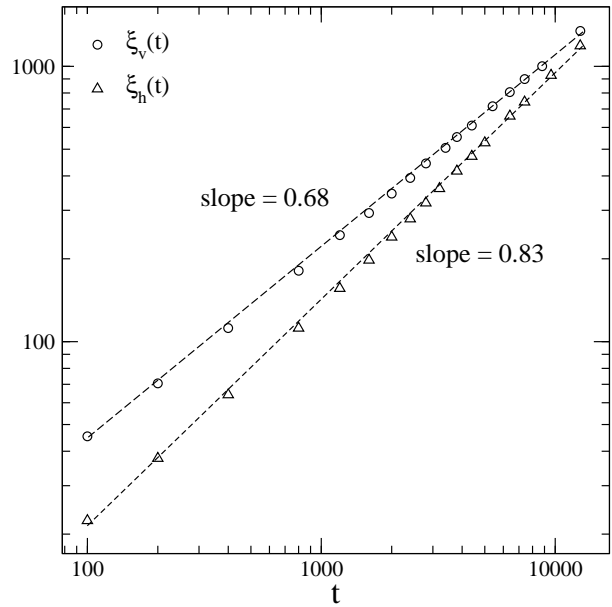}}}
\end{picture}
\caption{ The correlation lengths of the velocity $\xi_v(t)$ and height $\xi_h$(t) extracted from the scaling functions $\widetilde C_v(r,t)$ and $\widetilde C_h(r,t)$ are displayed with open circles and open triangles on a log-log scale, respectively. The dashed lines show power-law fits. }
\label{f4}
\end{figure*}

A comparison between the correlation lengths $\xi_v(t)$ and $\xi_h(t)$ obtained from  $C_v(r,t)$ and $C_h(r,t)$ is performed in Fig.~\ref{f4} with open circles and open triangles, respectively. Power-law behaviors are found for both of them, but the difference in the dynamic exponents, $1/z=0.68(1)$ and $1/z_h=0.83(1)$, reaches more than $20$ percent, confirming that the growth of velocity correlation length $\xi_v(t)$ is clearly distinguishable from that of the height-function one $\xi_h(t)$. We then estimate the dynamic exponent $z=1.47(2)$, consistent with that obtained from the cumulant $F_v(t)$ in Fig.~(\ref{f2}). So it makes sense to replace the correlation length $\xi_v(t)$ with the cumulant $F_v(t)$ for determining the transition point $H_c$ and critical exponents in Eqs.~(\ref{scaling 3}) and (\ref{scaling 4}).

\begin{figure*}[htb]
\centering
\epsfysize=8cm \epsfclipoff \fboxsep=0pt
\setlength{\unitlength}{1.cm}
\begin{picture}(9,8)(0,0)
\put(0.0,0.0){{\epsffile{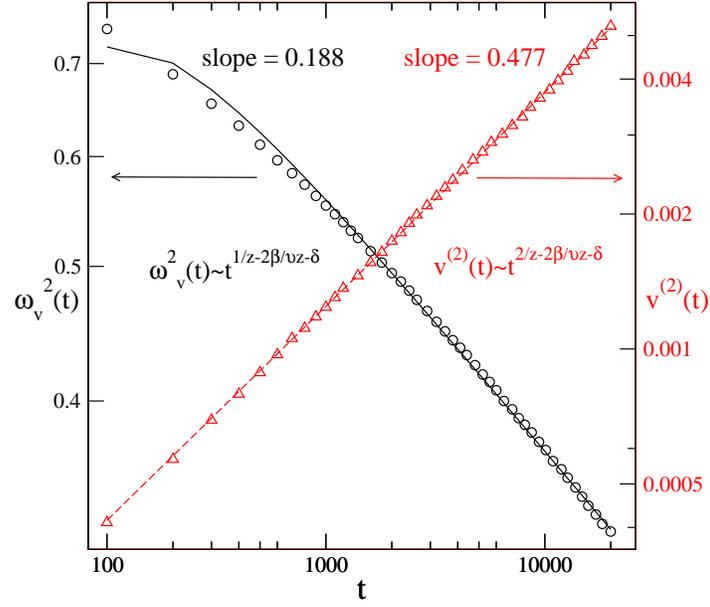}}}
\end{picture}
\caption{ The line susceptibility $\omega^2_v(t)$ and planar susceptibility $v^{(2)}(t)$ of the domain-wall velocity are plotted  with open circles and open triangles on a log-log scale, respectively. The dashed line represents a power-law fit, and the solid line shows a power-law fit with the correction. The right and left arrows indicate the $y$ coordinates of $\omega^2_v(t)$ and $v^{(2)}(t)$, respectively. }
\label{f5}
\end{figure*}

Furthermore, velocity fluctuations including the line susceptibility $\omega^2_v(t)$ and planar susceptibility $v^{(2)}(t)$ are investigated in Fig.~\ref{f5}. The former is expected to decay as a power-law form according to Eq.~(\ref{scaling 8}), although there exists a correction to scaling. Then one can roughly estimate the critical exponent $1/z-2\beta/\nu z-\delta=-0.188(3)$. From the scaling form in Eq.~(\ref{scaling 9}) and the critical behavior of $\widetilde C_v(r,t)$ mentioned in Fig.~\ref{f3}(a), one derives the formula $C_v(r,t)\sim \xi_v(t)r^{-2\beta/\nu-\delta z}$ when the distance is short $r < \xi_v(t)$. It indicates that the scaling behavior of the dynamic transition is significantly different from that of the equilibrium transition where $C(r,t)\sim r^{-2\beta/\nu}$ is independent of the time. In contrast, the planar susceptibility $v^{(2)}(t)$ follows a power-law increase. A direct measurement of the slope gives the exponent $2/z-2\beta/\nu z-\delta=0.477(5)$, yielding the overhang exponent $\delta=0.40(1)$, consistent with that obtained from $\widetilde C_v(r,t)$.

\begin{figure*}[htb]
\centering
\epsfysize=8cm \epsfclipoff \fboxsep=0pt
\setlength{\unitlength}{1.cm}
\begin{picture}(9,8)(0,0)
\put(0.0,0.0){{\epsffile{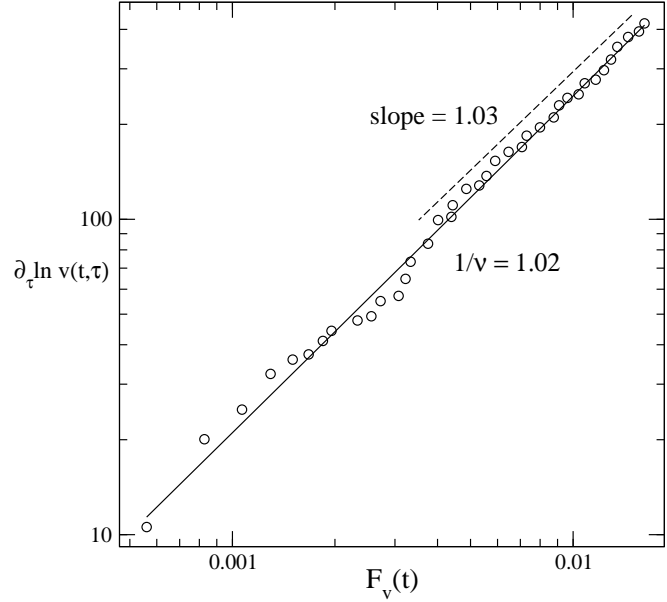}}}
\end{picture}
\caption{ The logarithmic derivative of $v(t,\tau)$ with respect to the transition deviation $\tau=|H-H_c|$ is displayed against the cumulant $F_v(t)$ on a log-log scale. The dashed line represents a power-law fit, and the solid line shows a power-law fit with the correction. }
\label{f6}
\end{figure*}

In Fig.~\ref{f6}, logarithmic derivative $\partial_{\tau}\ln v(t,\tau)$ is plotted as a function of the cumulant $F_v(t)$ at the critical point $H_c=1.4206$ with open circles. Quadratically interpolation of $v(t,\tau)$ is taken between the driving fields $H=1.420$ and $1.421$ for the convenience of calculations.  The exponent $1/\nu=1.03(3)$ is measured from the slope of the dashed line. After introducing a power-law correction to scaling in Eq.~(\ref{correction}), the fitting to numerical data extends to early times, and it yields the exponent $1/\nu=1.02(3)$. Accordingly, the velocity exponent $\beta=0.325(5)$ is calculated from $\beta/\nu=0.333(1)$ in Fig.~\ref{f1}, in agreement with experimental results \cite{alb21a}.

\begin{figure*}[htb]
\centering
\epsfysize=8cm \epsfclipoff \fboxsep=0pt
\setlength{\unitlength}{1.cm}
\begin{picture}(9,8)(0,0)
\put(0.0,0.0){{\epsffile{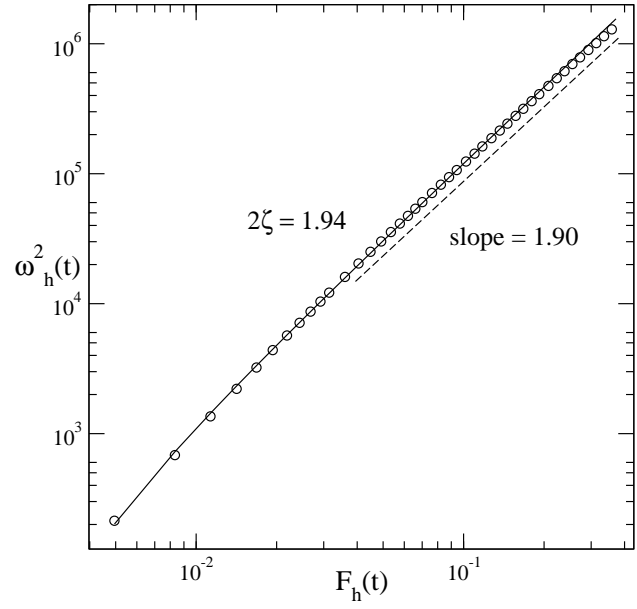}}}
\end{picture}
\caption{The roughness function $\omega^2_h(t)$ is plotted versus the susceptibility ratio $F_h(t)$ on a log-log scale. The dashed line represents a power-law fit, and the solid line shows a power-law fit with the correction. }
\label{f7}
\end{figure*}

Finally, the roughening process of the domain wall is studied for the depinning transition. In Fig.~\ref{f7}, the roughness function $\omega_h^2(t)$ is plotted against the ratio $F_h(t)$ on a log-log scale. The dashed line indicates a power-law behavior with the slope $1.90(2)$. The correction to scaling is also considered for extending the fitting to earlier times, which leads to the exponent $2\zeta=1.94(6)$. The roughness exponent $\zeta=0.97(3)\approx 1$ is thus obtained, showing a significant discrepancy with that of QEW equation.

\subsection{Influence of the disorder }
\begin{figure*}[htb]
\centering
\epsfysize=8cm \epsfclipoff \fboxsep=0pt
\setlength{\unitlength}{1.cm}
\begin{picture}(9,8)(0,0)
\put(0.0,0.0){{\epsffile{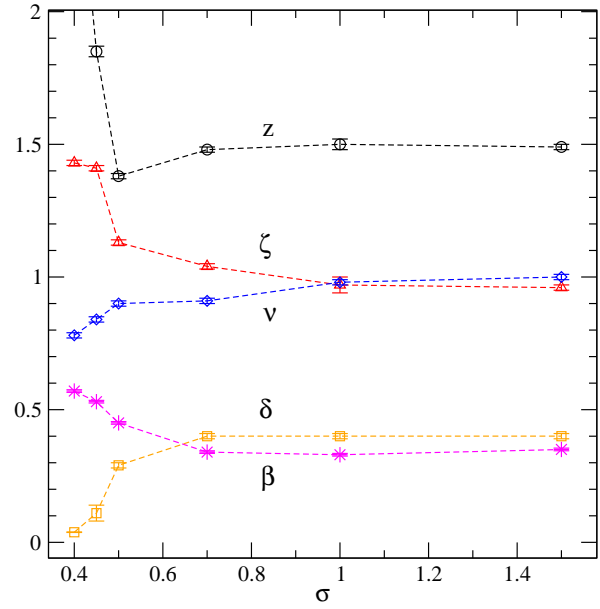}}}
\end{picture}
\caption{ Variations of the critical exponents are shown as a function of the disorder strength $\sigma$ on a linear scale. Open circles, open triangles, open diamonds, open squares, and stars correspond to the dynamic exponent $z$, roughness exponent $\zeta$, correlation-length exponent $\nu$, overhang exponent $\delta$, and velocity exponent $\beta$, respectively. Most of the error bars are smaller than the symbols. }
\label{f8}
\end{figure*}

In this subsection, the influence of the disorder on the depinning transition is studied by comprehensive simulations with the disorder strength ranging from $\sigma=0.2$ to $1.5$. The values of the critical exponents including the dynamic exponent $z$, roughness exponent $\zeta$, correlation-length exponent $\nu$, velocity exponent $\beta$, and overhang exponent $\delta$ are estimated from a variety of measurements mentioned before, and disorder-dependent behaviors are demonstrated in Fig.~\ref{f8}. The robustness of the exponents is affirmed for the disorder $\sigma>0.5$, yielding a new universality class:  $\beta=0.34(1), z=1.49(2), \zeta=0.97(5),\nu=0.96(4)$, and $\delta=0.40(1)$. It significantly differs from that of the QEW equation with depinning exponents $\beta = 0.245(6), z = 1.433(7), \zeta = 1.250(5)$, and $\nu = 1.333(7)$ which are also obtained from nonsteady dynamics simulations \cite{fer13}.

In addition, the hyperscaling relation $\nu=1/(2-\zeta)$ reflecting the statistical tilt symmetry (STS) is examined. Within the error bars, our results support this hyperscaling, showing the reliability of the universality class. Due to the existence of two distinct correlation lengths $\xi_v$ and $\xi_h$, another dynamic exponent $z_h=1.20(5)$ characterizing the roughness of the domain wall should also be taken into account, which is much smaller than the value of $z$. The scaling relation $\beta/\nu z + \zeta/z_h \approx 1$ is thus satisfied, refining the usual hyperscaling $\beta=\nu(z-\zeta)$. It shows that the roughening process of the domain wall is profoundly affected by the depinning transition.

\begin{figure*}[htb]
\centering
\epsfysize=8cm \epsfclipoff \fboxsep=0pt
\setlength{\unitlength}{1.cm}
\begin{picture}(9,8)(0,0)
\put(0.0,0.0){{\epsffile{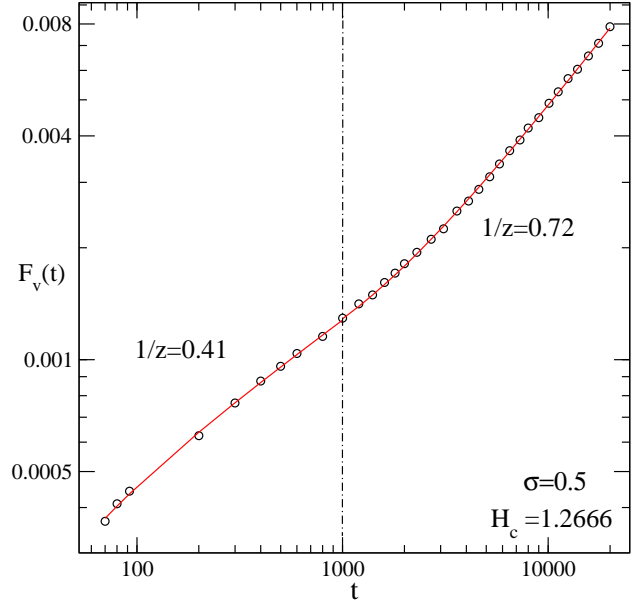}}}
\end{picture}
\caption{ At the medium disorder $\sigma=0.5$, the cumulant of the domain-wall velocity $F_v(t)$ is plotted at the transition field $H_c=1.2666$ on a log-log scale. The dash-dotted line indicates a two-stage kinetics with the values of the dynamic exponent $1/z\approx0.41$ and $0.72$, respectively. The solid line shows a power-law fit with the correction. }
\label{f9}
\end{figure*}

However, all critical exponents exhibit strong departures from those of the universality class when the disorder $\sigma \le 0.5$. In order to understand what happens, nonsteady critical behavior of the domain wall is analyzed subsequently for weaker disorders. At the disorder $\sigma=0.5$, the cumulant $F_v(t)$ is displayed in Fig.~\ref{f9} at the transition point $H_c=1.2666$ on a log-log scale.  A two-stage increase of the correlation length is found with the values of the dynamic exponent $1/z=0.41(1)$ and $0.72(1)$ at earlier and later times, respectively, quite different from that in Fig.~(\ref{f4}). It implies the existence of the crossover in the nonsteady relaxation. As reported in the literature \cite{fer13}, there are three different stages of the dynamics at the depinning transition, which are microscopic, mesoscopic, and macroscopic, respectively. They are separated by two characteristic time scales $t_{\rm mic}$ and $t_{\rm mes}$, and critical exponents are universal only in the macroscopic critical regime. In this work, all the results are measured at $t \ge t_0 >t_{\rm mic}=20$ MCS where $t_0=100$ MCS, so the effect of microscopic time regime is negligible. The nontrivial long crossover i.e., mesoscopic time regime, appears in the case of weak disorder, and the time scale $t_{\rm mes}$ diverges in the limit $\sigma \rightarrow 0$. Taking $\sigma=0.45$ as an example, $t_{\rm mes}$ is larger than  $t_{\rm max}=20000$ MCS, showing the simulations we perform are always in the mesoscopic time regime, resulting in nonuniversal values of the exponents. In contrast, $t_{\rm mes}\approx 200$ MCS is obtained at a strong disorder $\sigma=1.0$ as shown in Fig.~(\ref{f2}). The mesoscopic time regime thus has little influence on the critical behaviors which can be fairly described as a power-law correction to scaling.

\begin{figure*}[htb]
\centering
\epsfysize=8cm \epsfclipoff \fboxsep=0pt
\setlength{\unitlength}{1.cm}
\begin{picture}(9,8)(0,0)
\put(0.0,0.0){{\epsffile{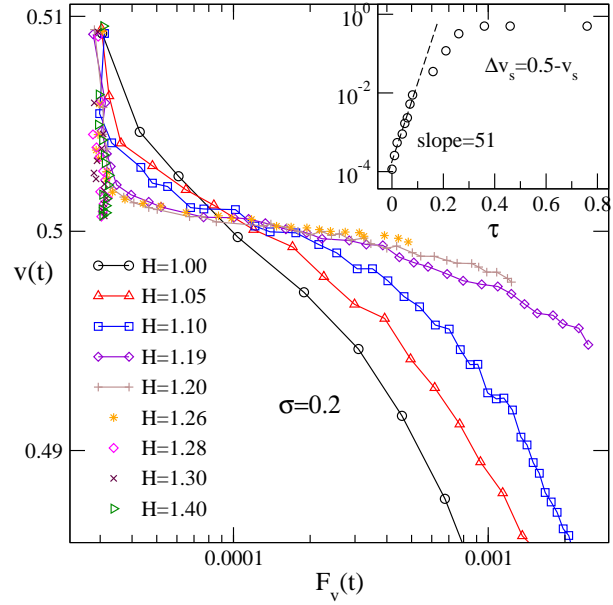}}}
\end{picture}
\caption{ For the weak disorder case with $\sigma=0.2$, the domain-wall velocity $v(t)$ is displayed as a function of the cumulant $F_v$(t) for different driving fields $H$ on a log-log scale. In the inset, the shift of the steady-state velocity $\Delta v_s=0.5-v_s$ is plotted against the transition deviation $\tau=|H-H_c|$ on a log-linear scale. The dashed line represents an exponential fit. }
\label{f10}
\end{figure*}

Furthermore, the case far away from the universality regime is investigated at $\sigma=0.2$, and the results are presented in Fig.~\ref{f10}. The order parameter $v(t)$ versus the cumulant $F_v(t)$ is displayed for different driving fields $H$. Obviously distinct from that in Fig.~\ref{f1}, no power-law behavior is detected, and the curves are almost overlapped for the fields $H>H_c=1.26$ wherein the correlation length represented by the cumulant $F_v(t)$ drops rapidly with $H$. It indicates that the transition may be not of second order. In the inset, the shift of the steady velocity $\Delta v_s=0.5-v_s$ is plotted against the transition deviation $\tau=H-H_c$. An exponential fit is carried out with the dashed line, and the value of the slope $51$ is quite large, indicating that there is a sharp decline of the velocity when the driving field $H$ is close to the transition point $H_c$. It suggests that the depinning transition is discontinuous when the disorder strength is low enough. Moreover, a crossover from the first-order to second-order phase transition is observed, just as that in the model with uniformly distributed disorder. \cite{zho16}.

\begin{figure*}[htb]
\centering
\epsfysize=7cm \epsfclipoff \fboxsep=0pt
\setlength{\unitlength}{1.cm}
\begin{picture}(10,7)(0,0)
\put(-2.0, 0.0){{\epsffile{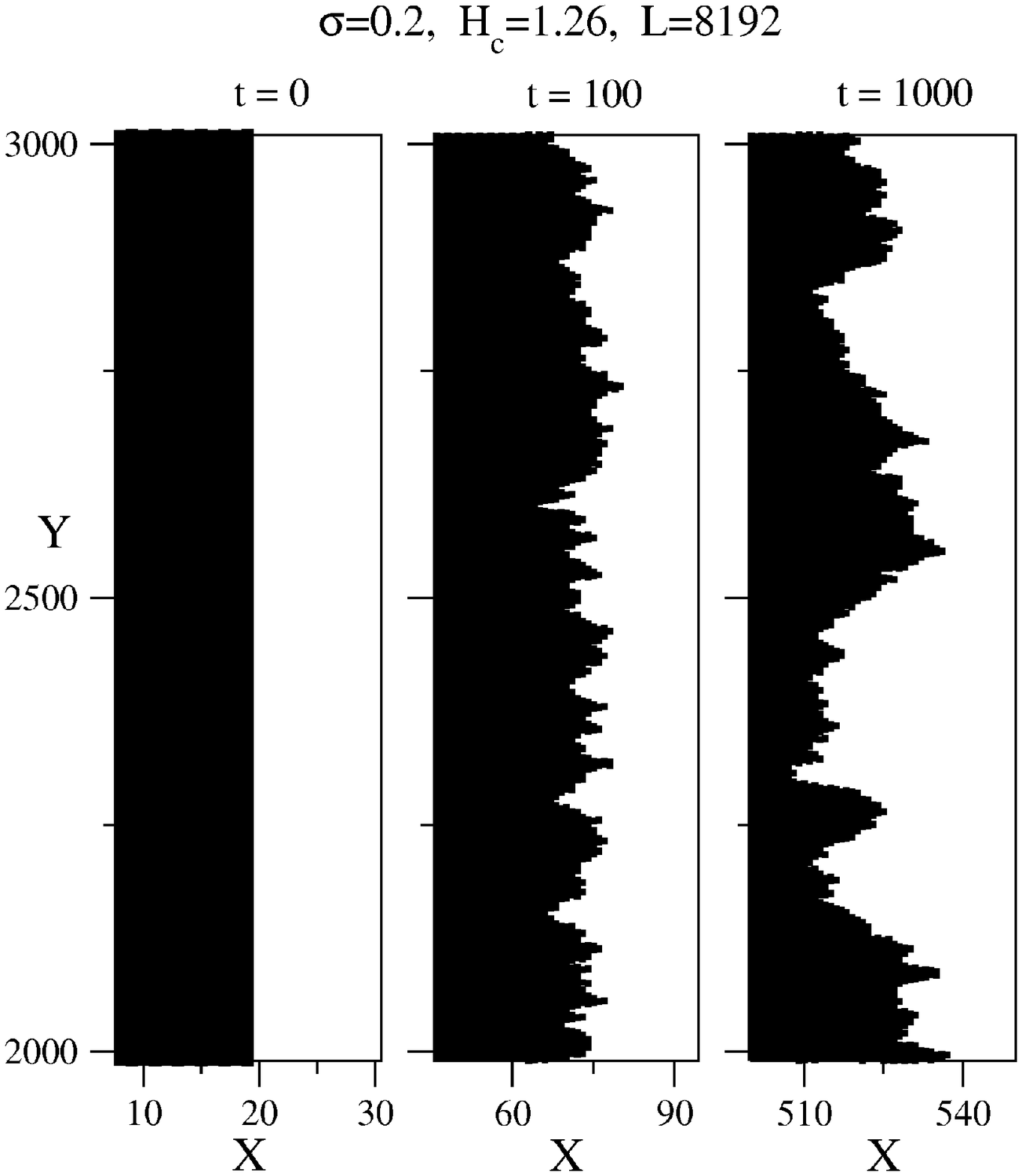}}}\epsfysize=7cm
\put(4.7, 0.0){{\epsffile{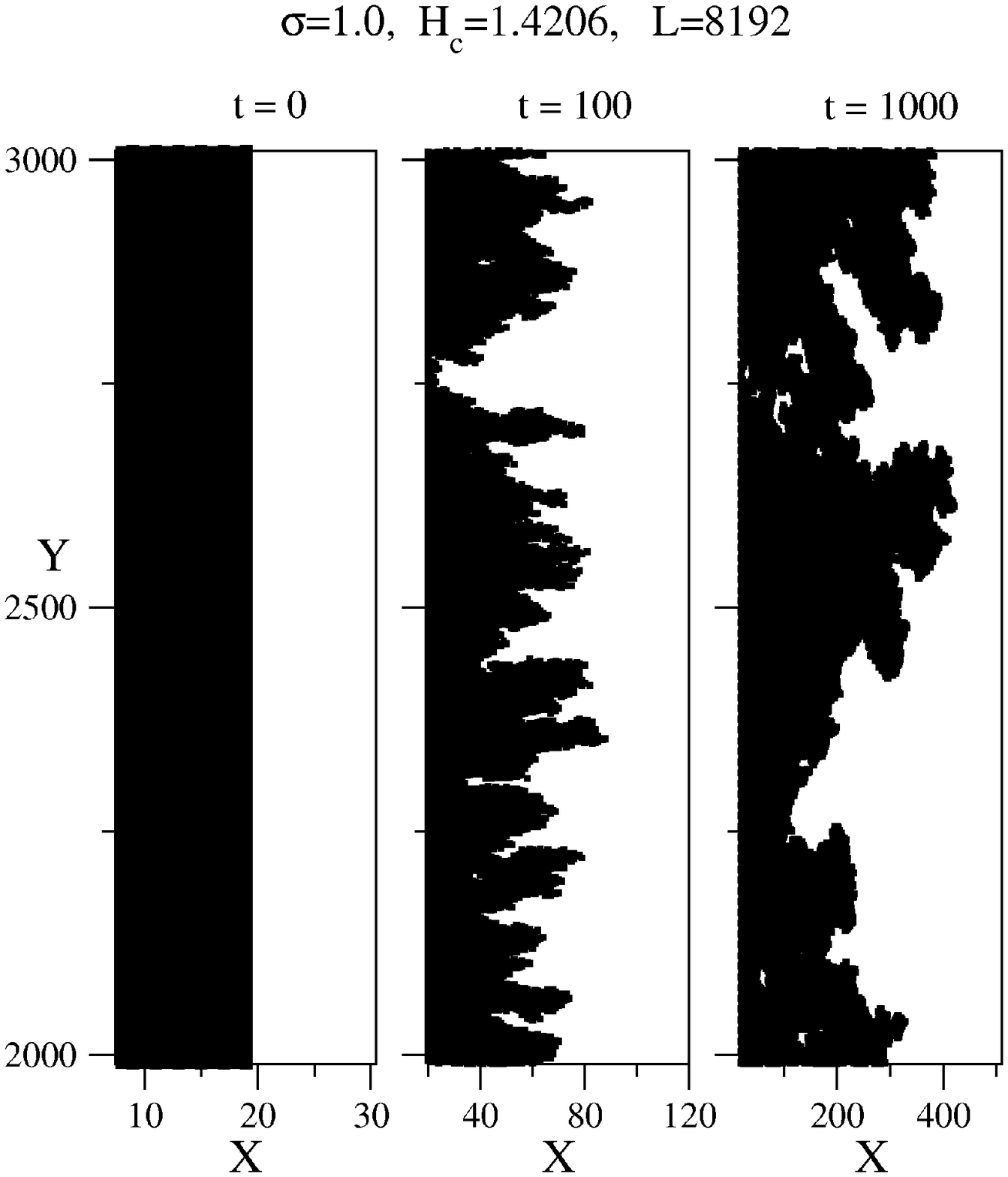}}}
\end{picture}

\hspace{0.0cm}\footnotesize{(a)}\hspace{6.2cm}\footnotesize{(b)}
\caption{Time evolution of the spin configurations under the Gaussian distribution of the random fields is shown at the disorder strengths $\sigma=0.2$ in (a) and $\sigma=1.0$ in (b). The black and white correspond to the spin up $S_i=1$ and down $S_i=-1$, respectively. The critical driving fields $H_c=1.26$ and $1.4206$ are set. }
\label{f11}
\end{figure*}

Time evolution of the spin configurations in the weak-disorder regime ($\sigma=0.2$) and strong-disorder regime ($\sigma=1.0$) is displayed in Fig.~\ref{f11} to provide a better understanding of the depinning transition. The black and white squares correspond to $S_i=\pm 1$, respectively. In the subfigure (a), the domain wall moves fast but roughens slightly, with the width much smaller than the lattice size. The reason is that the correlation length does not diverge at the first-order phase transition.  Conversely, the width of the domain wall in the subfigure (b) increases sharply with time, and finally diverges if the lattice size is infinite, signifying the system exhibits a divergent correlation length at the second-order phase transition. Moreover, complicated spin structures such as overhangs and islands are found nearby the domain interface at the time $t = 1000$ MCS, which differ considerably from those in the subfigure(a). It indicates overhangs and islands have an essential effect on the nonsteady critical dynamics when the disorder is strong. However, the effect becomes negligible in the weak-disorder regime, agreeing well with the behavior of the overhang exponent $\delta$ in Fig.~\ref{f8}.

\begin{figure*}[htb]
\centering
\epsfysize=8cm \epsfclipoff \fboxsep=0pt
\setlength{\unitlength}{1.cm}
\begin{picture}(9,8)(0,0)
\put(0.0,0.0){{\epsffile{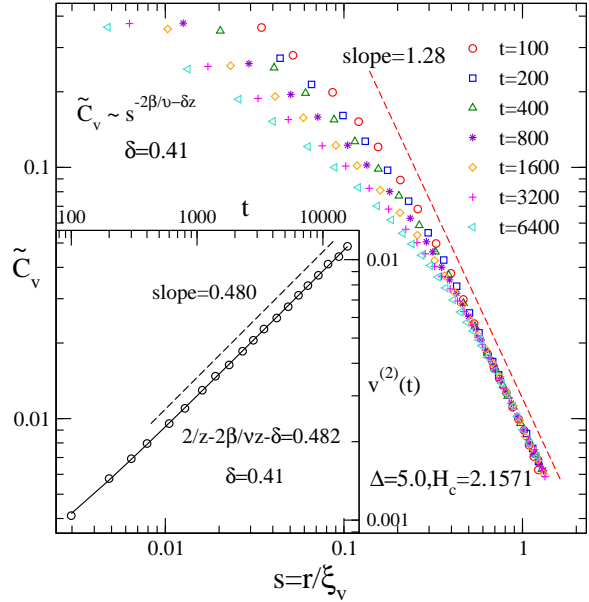}}}
\end{picture}
\caption{In the model with the disorder uniformly distributed within an interval $[-\Delta, \Delta]$, the scaling function of the velocity correlation $\widetilde{C}_v(s)$ is shown with respect to $s=r/\xi_v (t)$ for different times on a log-log scale. The strength of the disorder $\Delta=5.0$ is taken as an example in the strong-disorder regime. Data collapse is demonstrated at the critical filed $H_c=2.1571$, and the overhang exponent $\delta=0.41(2)$ is estimated. In the inset, the global velocity fluctuation $v^{(2)}(t)$ is presented. The dashed lines represent power-law fits, and the solid line shows a power-law fit with the correction. }
\label{f12}
\end{figure*}

For comparison, additional simulations have been conducted in the random-field Ising model with uniform disorder ranging from $[-\Delta, \Delta]$. The disorder strength $\Delta=5.0$ is set as an instance in the strong-disorder regime. At the critical point $H_c=2.1571$,  nonequilibrium dynamics of the depinning transition has been investigated with the same lattice size $L=8192$ and time scale $t_{\rm max}=20000$ MCS. In Fig.~\ref{f12}, the scaling function of the velocity correlation $\widetilde{C}_v(s)$ is plotted for different times $t$ on a log-log scale. In the same way, data collapse onto a single curve is obtained, and the overhang exponent $\delta=0.41(2)$ is estimated from the slope $1.28$, which is almost identical to that in the inset of Fig.~\ref{f3}(a). Besides, global velocity fluctuation $v^{(2)}(t)$ has been analyzed in the inset, and the value $\delta=0.41(1)$ again corroborates the notion that the contribution of overhangs and islands in the model with uniform distribution converges to that in the model with Gaussian one at strong disorder. Further studies on critical exponents $\beta, \nu, z$, and $\zeta$ show that the transitions in both models belong to the same universality class.

\section{Conclusion}\label{sec:conclusion}
The dynamic depinning transition in two-dimensional Gaussian random-field Ising model has been systematically investigated with large-scale simulations up to $L=8192$ and $t_{\rm max}=20000$ MCS based on the extended Monte Carlo algorithm. The nonsteady dynamics at the dynamic transition has been carefully examined, and dynamic scaling behaviors of the order parameter and its fluctuation and correlation have been identified as well as the roughness and height correlation of the domain wall. The transition field, static and dynamic exponents have been accurately determined by the STCD analysis, and the influence of the disorder has also been revealed. Unlike the equilibrium transition, two distinguished growth processes of spatial correlation lengths $\xi_v(t)$ and $\xi_h(t)$ have been found for the dynamic depinning transition. Besides, a ``new'' dynamic exponent $\delta$ related to the effect of overhangs and islands has been extracted from the dynamic scaling behaviors of the velocity fluctuation and correlation.

In particular, we have uncovered the universality class of the depinning transition in Ising-type lattice models: $\beta=1/3,\nu=1,\zeta=1, \delta=2/5$, and $z=3/2$, which significantly differs from that of the QEW equation. Both the hyperscaling relations $\nu=1/(2-\zeta)$ and $\beta/\nu z+\zeta/z_h=1$ hold, lending strong support to the reliability of the universality class. Moreover, our results agree well with those of recent experiments $\beta=0.30(3)$ and $z=1.5(2)$  in ferrimagnetic GdFeCo thin films \cite{alb21a}, dislocation depinning simulations $\beta=0.30(5), \nu=1.05(5)$, and $\zeta=0.96(2)$ in dispersion-strengthened steels \cite{bak08}, and molecular dynamics simulations $\beta=0.29(3)$ and $\nu=1.04(4)$ in two-dimensional vortex lattices \cite{sca12}. What they have in common is that detailed microscopic structures and interactions are contained which are lacking in the QEW equation.

In the weak disorder regime, however, the crossover from the second-order phase transition to the first-order one takes place due to the appearance of the mesoscopic time regime, yielding the exponents deviating dramatically from those of the universality class. Interestingly, both models with Gaussian and uniformly distributed disorder feature such a crossover in the phase diagram, despite differing in the transition boundary and exponents. They converge to the same universality class when subjected to strong-disorder regime.
It suggests that the disorder strength is relevant for the criticality of the depinning transition, as it is claimed in the recent numerical work based on the LLG equation \cite{ska22}.

{\bf Acknowledgements:} This work was supported by National Natural Science Foundation of China under Grant Nos. $11875120$.

\bibliography{zheng,domain}
\bibliographystyle{apsrev4-1}

\end{document}